\documentclass{article}
\usepackage{graphicx}  % standard LaTeX graphics tool;
\usepackage{amsmath}   % For getting proper math eqns
\usepackage{amssymb}   % Used for getting various symbols eg. gtrsim, lesssim etc.
\usepackage{bm} % For bold math (esp of lower greek letters)
\usepackage{dcolumn}% Align table columns on the decimal point
\usepackage{color}
\usepackage{mathrsfs}
\usepackage{amsfonts}
\usepackage{varioref}
\usepackage{slashed}
\usepackage{amsmath}
\usepackage{cite}
\usepackage{authblk} % for adding affliliations
\RequirePackage[colorlinks,citecolor=blue,urlcolor=magenta,linkcolor=blue]{hyperref}
\usepackage{comment}
\def\be{\begin{equation}}
\def\ee{\end{equation}}

\labelformat{section}{Section #1} 
\labelformat{subsection}{Section #1} 
\labelformat{subsubsection}{Section #1}
\labelformat{subsubsubsection}{Section #1}
\labelformat{equation}{Eq.~(#1)} 
\labelformat{figure}{Fig.~#1} 
\labelformat{subfigure}{Fig.~\thefigure#1} 
\labelformat{table}{Tab.~#1} 
\labelformat{appendix}{Appendix #1}

\title{\bf Backreaction inclusive Schwinger effect in flat and de Sitter spacetimes via a self-consistent Maxwell-Schr\"odinger semiclassical dynamics} %\\ \textcolor{red}{Beyond Adiabaticity: Real-Time Pair Production and Field Dynamics in Curved Spacetime}}
\author[1,2]{Shagun Kaushal\thanks{shagun123@iitd.ac.in, shagun.kaushal@vit.ac.in}}
\author[1]{Suprit Singh\thanks{suprit@iitd.ac.in}}

\affil[1]{Department of Physics, Indian Institute of Technology Delhi\\
Hauz Khas, New Delhi, India 110016}
\affil[2]{Department of Physics, Vellore Institute of Technology\\
Vellore, Tamil Nadu, India 632007 \footnote{Current affiliation}}

\date{\today}  %% This command will suppress printing the date. 

\begin{document}
\maketitle
\begin{abstract}
\noindent
We employ a self-consistent framework to study the backreaction effects of particle creation in the coupled semiclassical dynamics of a quantum complex scalar field and a classical electric field in both (1 + 1)- and (1 + 3)-dimensional Minkowski and de Sitter spacetimes. Using a general Gaussian-state formalism in the Schrödinger picture, we solve the resulting nonlinear equations with Gaussian initial data, obtaining a self-consistent semiclassical evolution that incorporates nonperturbative backreaction. We compute the time-dependent instantaneous particle content, current density, and electric field, defined through instantaneous eigenstates of the field modes. Comparing scenarios with and without backreaction, we find that backreaction strongly modifies the electric field and current, producing immediate plasma-like oscillations and driving pronounced oscillations in the instantaneous mode occupations through nonadiabatic squeezing and quantum interference. These oscillations do not imply additional irreversible particle production—the time-averaged particle number remains essentially constant—but they reveal the rich nonperturbative real-time dynamics captured by our self-consistent semiclassical approach across dimensions and in both Minkowski and de Sitter backgrounds.
\end{abstract}
%\tableofcontents
%%%%%%%%%%%
\section{Introduction}\label{S1}

Schwinger effect \cite{Schwinger1, Schwinger2, Schwinger3, Schwinger4} is one of the predictions of quantum electrodynamics, where we have vacuum polarization as well as the decay of the vacuum into charged particle pairs under strong external electric fields \cite{grib}. Pair creation in the Schwinger effect is a well studied phenomenon theoretically, but still evades experimental confirmation, as the pair creation rate is exponentially small for a homogeneous electric field configuration. Various efforts in this direction are underway, either by considering time-varying electric fields via high-intensity lasers, or dynamically assisted mechanisms via modulations \cite{Dunne:2010zz, Dunne:2014qda, HLI, Schützhold2008,Popov:2001ak, Allor:2007ei, Schmitt:2022pkd}. The Schwinger mechanism gains an additional importance in the case of inflationary physics\cite{Parker:1968mv, Parker:1969au, Parker:1971pt, Parker:2012at} and in charged black holes where it works in tandem with the gravitational particle production. As such the effect has been broadly studied for a homogeneous electric field configuration in flat and curved spacetimes (including de Sitter (dS), anti-de Sitter (AdS), the Rindler, and many more) as well as in different spatial dimensions \cite{Kim:2016xvg, Srinivasan:1998fk, vilenkin, Sharma:2017ivh, Xue:2017cex, yoko:2016tty, Xue:2017ecx, SHG, SHU, out_def, Gabriel:1999yz, Kobayashi:2014zza}. Recently, refs.~\cite{Hebenstreit:2011pm, Hu:2022ouk} have considered the Schwinger effect in inhomogeneous electric fields using the Dirac-Heisenberg-Wigner formalism, albeit in (1+1) dimensions. 

%However, the threshold electric field required for this pair creation to occur is estimated as $10^{18} Vm^{-1}$, which is far too high to be achieved with nowadays technology. The next generation of high-energy lasers, such as the Extreme Light Infrastructure (ELI) \cite{HLI}, will aim to achieve the critical value of the electric field required for detecting the Schwinger effect. 

Most studies of the Schwinger effect have assumed no backreaction: the classical electric field is treated as an external background with its own independent dynamics, unaffected by the quantum subsystem and in particular by the particles it produces. In realistic situations, however, the induced current from the produced pairs will inevitably modify the progenitor field, so incorporating backreaction is essential for a self-consistent description. Backreaction of Schwinger pairs on a time-varying but spatially homogeneous electric field has been examined in ~\cite{BRSE1, BRSE2, BRSE3, BRSE4, BRSE5, BRSE6, BRSE7, BRSE8, Gavrilov:2012jk, Copinger:2024pai, Lozanov:2018kpk, Maleknejad:2019hdr}, but this represents only a partial treatment as one expects the development of inhomogeneities in the background field due to pair nucleation. A simple depiction of the expected spatial effects is given in ref.~\cite{Nuclear Physics B297 (1988) 787-836}, where instanton techniques showed that the electric field splits into $E_{\mathrm{i}}$ and $E_{\mathrm{o}}$ on the two sides of a nucleated particle’s worldline, with a jump discontinuity at the particle trajectory. A rigorous field-theoretic computation of such spatial effects is still lacking and requires a backreaction-inclusive semiclassical dynamics of the combined (electric field + quantum field) system, since a fully quantum-mechanical treatment of the entire system is presently intractable. In the present work we restrict to spatially homogeneous but time-dependent electric fields, implementing a fully self-consistent treatment of their temporal backreaction. Our formalism, however, is general enough to be extended to spatially varying fields in future work. The question now is how to prescribe a consistent “semiclassical” or hybrid evolution for the interacting classical and quantum dynamical variables. 

Semiclassical effects are usually extracted from the “in–out” effective action \cite{MRBrown}, obtained by integrating out the quantum degrees of freedom in the global path integral while keeping the classical background fixed. The resulting effective action is generally complex, where the real part corresponds to the vacuum polarisation and the imaginary part corresponds to the vacuum decay or the pair production rate. However, this approach assumes a fixed classical background and also relies on the existence of stationary ``in" and ``out" states in the asymptotic past and future. This framework does not work in the case of an explicitly time-dependent spacetime background, such as FLRW spacetime, which lacks Killing vectors in time. Moreover, incorporating the imaginary part of the effective action into a consistent semiclassical evolution is problematic because adding it directly renders the real equations complex. When the imaginary part is much smaller than the real part, one usually neglects it and works only with the real part \cite{DeWitt paper}, but this prescription becomes questionable when particle production is significant. Various other semiclassical proposals with backreaction effects on classical systems can be found in refs.~\cite{SC1, SC2, SC3, SC4, WBR1, WBR2, WBR3, WBR4, WBR5, WBR6, WBR7, WBR8, WBR9}. 

To overcome these limitations, we adopt the local canonical prescription developed in \cite{Husain:2021rnf, Husain:2018fzg}. This framework (i), being local, allows transparent and self-consistent dynamics incorporating backreaction, and (ii) permits a well-posed Cauchy problem formulated in terms of coupled differential equations that can be solved with appropriate initial and boundary data—analytically where possible and numerically otherwise. Next, we briefly summarise this prescription and its salient features.

To implement this local canonical prescription we explicitly split the system into two degrees of freedom, \(C\) and \(Q\), where \(Q\) is treated quantum mechanically and \(C\) as a classical variable. These subsystems interact through a Hamiltonian \(H(C,Q)\). After fixing any gauge choices we obtain a physical Hamiltonian which we assume to be separable as
\[
H(C,Q)=H_{1}(C)+H_{2}(Q,C;\alpha),
\]
where the second term incorporates any interactions via coupling constants \(\alpha\). In some cases \(H_{2}\) can be further decomposed into a free Hamiltonian plus an interaction term, but in others—especially in curved spacetime—this is not possible and the notion of in/out states breaks down. Consequently, the usual canonical approach based on Bogoliubov coefficients requires adiabatic approximations and still does not handle the full problem.

To avoid this limitation we evolve \(Q\) quantum mechanically according to a time-dependent Schrödinger equation
\begin{equation}
\label{quant}
\hat{H}_{2}(Q,C)\,\psi(Q,t)=i\hbar\frac{\partial\psi}{\partial t}(Q,t),
\end{equation}
where \(C\) enters as a \(c\)-number. The classical degree of freedom evolves via
\begin{equation}
\label{class}
\dot{C}=\{C,H_{\text{eff}}\},\qquad 
H_{\text{eff}}=H_{1}(C)+\langle\psi|\hat{H}_{2}(Q,C)|\psi\rangle,
\end{equation}
which is a Poisson bracket of \(C\) with an effective Hamiltonian. Together, these two equations define a \emph{self-consistent semiclassical evolution} to be solved with initial data \(\{\psi_{0},C_{0}\}\). All observables can then be computed at each time step without relying on the existence of adiabatic vacua or asymptotic in/out states, making the formalism well-suited to explicitly time-dependent scenarios without needing any adiabatic approximation.

% \begin{figure}
%      \centering
%      \includegraphics[width=0.3\linewidth]{Block diagram.jpg}
%      \caption{Flowchart illustrating the semiclassical backreaction framework. The quantum field evolves via the time-dependent Schrödinger equation, leading to particle production. The resulting current modifies the classical electric field, which in turn influences the quantum field state, forming a self-consistent feedback loop.}
%     \label{fig:semiclassical_flow}
%  \end{figure}

%Schwinger mechanism forms an apt testbed for the above framework. In this paper, we consider scalar quantum electrodynamics where a quantum charged scalar field $\phi(t,\bf x)$ is subjected to an external classical electric field, ${\bf E}(t,{\bf x})$ in $(1+1)$ and $(1+3)$ dimensional Minkowski and de Sitter spacetimes. Our focus is on studying this setup in both flat and cosmological de Sitter spacetimes, with a particular interest in analyzing the effects of backreaction on pair production, the electric field, and the resulting current density. By the above prescription, we treat the quantum field evolution using a time-dependent Schr\"odinger equation (TDSE) for each mode, where the classical electric field enters as a $c$-number, while the classical background evolves through Hamiltonian dynamics using Poisson brackets with an effective Hamiltonian containing quantum expectation values. 

Having established the general framework, we now apply it to the Schwinger mechanism, which provides an apt testbed. In this paper, we consider scalar quantum electrodynamics, where a quantum charged scalar field \(\phi(t,\mathbf{x})\) interacts with an external classical electric field \(\mathbf{E}(t,\mathbf{x})\) in \((1+1)\)- and \((1+3)\)-dimensional Minkowski and de Sitter spacetimes. We aim to study this setup in both flat and cosmological backgrounds with particular emphasis on how backreaction affects pair production, the electric field, and the resulting current density.  Within our prescription, the quantum field evolution is governed, for each mode, by a time-dependent Schrödinger equation (TDSE) in which the classical electric field enters as a \(c\)-number. The classical background evolves separately through Hamiltonian dynamics via Poisson brackets with an effective Hamiltonian that contains the appropriate quantum expectation values.

Having outlined our framework and the system under study, the remainder of the paper is organized as follows. In \ref{sec:canonical}, we examine a canonical approach for coupling a complex scalar quantum field to a classical electric field in the Minkowski spacetime and calculate the particle number in each mode as a function of time, neglecting backreaction effects. Next, we introduce the general concept of backreaction from particles created in the presence of a background electric field in the Minkowski spacetime in \ref{Schwinger effect with backreaction}. We examine the effect of backreaction on the electric field, the current density, and the particles created. Further, in \ref{Schwinger effect in the cosmological de Sitter spacetime} and \ref{Schwinger effect with backreaction in de Sitter spacetime} we extend this analysis for the cosmological de Sitter spacetime. Finally, we summarise the paper in \ref{Conclusion and outlook}. Throughout this paper, we set $c = 1 =\hbar$.

\section{A canonical approach to the Schwinger effect}
\label{sec:canonical}

\label{A}
We begin by considering a complex scalar field in the presence of a background electric field in $(1+1)$-Minkowski spacetime and revisit the Schwinger pair production using the canonical approach (appropriately extending the formalism in \cite{Mahajan:2007qc, Mahajan:2007qg}). 
The Hamiltonian density of a complex scalar field $\phi(t,\bf x)$ of mass $m$ coupled to an external gauge field $A_\mu(x)= (0,A_1(t))$ in the Minkowski spacetime is
\begin{equation}
    \label{Hamiltoniandensity}
    \mathcal{H} = \frac{E^2(t)}{2}+\frac{1}{2}\Big(\Pi^\dagger \Pi+(\partial_1 - i q A_1(t))\phi^\dagger (\partial_1+i q A_1(t))\phi + m^2 \phi^\dagger \phi\Big)   
\end{equation}
where $\Pi$ and $E$ are the conjugate momenta associated with the complex scalar and gauge fields, respectively. 
Choosing the ansatz for the external gauge field as \( A_\mu = (0, A_1(t)) \) gives a nonzero electric field  
\begin{equation}
E(t) = -\frac{\partial A_1(t)}{\partial t}
\end{equation}
The form of \( A_1(t) \) determines whether the electric field is constant or time-dependent. In our analysis, we consider an initial seed electric field \( E = E_0 \) at \( t = 0 \), which imposes the condition  
\begin{equation}
E_0 = -\left. \frac{\partial A_1(t)}{\partial t} \right|_{t=0}
\end{equation}
This setup allows for different choices of \( A_1(t) \) depending on the desired time evolution of the electric field.

The total Hamiltonian is given by
\begin{equation}
    \label{TH}
    H=\int dx\; \mathcal{H}=\int dx \Bigg[\frac{E^2(t)}{2}+\frac{1}{2}\Big(\Pi^\dagger \Pi+(\partial_1 - i q A_1(t))\phi^\dagger (\partial_1+i q A_1(t))\phi + m^2 \phi^\dagger \phi\Big)  \Bigg]=H_1+H_2
\end{equation}
where the separable components are defined as follows: 
\begin{eqnarray}
\label{CQ}
    H_1 &=&\frac{1}{2}\int dx E^2(t)\\
    H_2 &=&\frac{1}{2} \int dx\Big(\Pi^\dagger \Pi+(\partial_1 - i q A_1(t))\phi^\dagger (\partial_1+i q A_1(t))\phi + m^2 \phi^\dagger \phi\Big) 
\end{eqnarray}
Here, $H_1$ corresponds to the completely classical part of the Hamiltonian, while $H_2$ represents the quantum part in which the classical electric field enters as a $c$-numbered field. In Fourier space, the Hamiltonian $H_2$ becomes

\begin{equation}
    \label{HamiltonianMomentum}
   H_2= \frac{1}{2}\int_{-\infty}^{\infty} \frac{d k_1 }{2 \pi} \Bigg[\Pi^\dagger_{k_1} \Pi_{k_1}+(k_1+q A_1(t))^2\phi_{k_1}^\dagger \phi_{k_1}+m^2\phi^\dagger_{k_1}\phi_{k_1} \Bigg] = H_{2}(k_1) \oplus H_{2}(-k_1)
\end{equation}
where \begin{eqnarray}
    H_{2}(k_1) = \frac{1}{2} \int_{0}^{\infty} \frac{d k_1 }{2 \pi} \Bigg[|\Pi_{k_1}|^2+(k_1+q A_1(t))^2|\phi_{k_1}|^2+m^2|\phi_{k_1}|^2 \Bigg]=\int_{0}^{\infty}\frac{dk_1}{2\pi}\hat{h}_{k_1}\\
     H_{2}(-k_1) = \frac{1}{2} \int_{0}^{\infty} \frac{d k_1 }{2 \pi} \Bigg[|\Pi_{-k_1}|^2+(|k_1|-q A_1(t))^2|\phi_{-k_1}|^2+m^2|\phi_{-k_1}|^2 \Bigg]=\int_{0}^{\infty}\frac{dk_1}{2\pi}\hat{h}_{-k_1}
\end{eqnarray}
here $\Pi_{k_1}$ (or $\Pi_{-k_1}$) and $\phi_{k_1}$ or ($\phi_{-k_1}$) are the Fourier transforms of conjugate momentum and complex scalar field. 

The quantum–classical framework is established by quantizing the scalar field and describing its dynamics using a time-dependent Schrödinger equation in which the electric field enters as a $c$-number. Being a field theory, we have a TDSE for each mode of the complex scalar field evolving independently. The state corresponding to \ref{HamiltonianMomentum} is

\begin{equation}
    \label{psidef}
    \psi = \prod_{k_1,-k_1} ( \psi_{k_1} \otimes \psi_{-k_1} )
\end{equation}
Given that there is no mode mixing, we will focus on a single mode for the remainder of the paper. The time-dependent Schr\"odinger equation (TDSE) for a  bipartite mode is
\begin{equation}
    \label{tdse1dS}
    i \frac{\partial}{\partial t} (\psi_{k_1} \otimes \psi_{-k_1})= (\hat{h}_{k_1}\otimes \hat{I}_{-k_1} \oplus \hat{I}_{k_1}\otimes \hat{h}_{-k_1} )(\psi_{k_1} \otimes \psi_{-k_1})
\end{equation}
Here $\hat{h}_{k_1}$ and $\hat{h}_{-k_1}$ are given by
\begin{eqnarray}
  \hat{h}_{k_1} = \frac{1}{2} \Big(|\Pi_{k_1}|^2+(k_1+q A_1(t))^2|\phi_{k_1}|^2+m^2|\phi_{k_1}|^2\Big) \\
   \hat{h}_{-k_1} =  \frac{1}{2}\Big(|\Pi_{-k_1}|^2+(|k_1|-q A_1(t))^2|\phi_{-k_1}|^2+m^2|\phi_{-k_1}|^2 \Big)
\end{eqnarray}
On decoupling, above equation, we will have two TDSE for $\psi_{k_1}$ and $\psi_{-k_1}$, given as
\begin{eqnarray}
    \label{tdse}
    i\frac{\partial}{\partial t}\psi_{k_1}(\phi_{k_1},A_1(t),t)=\hat{h}_{k_1}\psi_{k_1}(\phi_{k_1},A_1(t),t)\\ 
     \label{tdse2}
    i\frac{\partial}{\partial t}\psi_{-k_1}(\phi_{-k_1},A_1(t),t)=\hat{h}_{-k_1}\psi_{-k_1}(\phi_{-k_1},A_1(t),t)
    \end{eqnarray}

We solve \ref{tdse} using a form-invariant Gaussian ansatz of wavefunction $\psi_{k_1}$ given by
\begin{equation}
    \label{psi}
    \psi_{k_1}(\phi_{k_1},A_1(t),t)=\beta_{k_1}(t) \text{exp}[-\alpha_{k_1}(t)|\phi_{k_1}|^2]
\end{equation}
where $\alpha_{k_1}$ and $\beta_{k_1}$ can be complex in general, and  on normalizing, we have
\begin{equation}
    \label{alphabeta}
    |\beta_{k_1}|^2=\sqrt{\frac{2 \text{Re}(\alpha_{k_1})}{\pi}}
\end{equation}
Likewise, \ref{tdse2} can be solved by considering 
$\psi_{-k_1}(\phi_{-k_1},A_1(t),t)=\beta_{-k_1}(t) \text{exp}[-\alpha_{-k_1}(t) |\phi_{-k_1}|^2 ]$.
     
On substituting the above ansatz in \ref{tdse}, equations of motion for $\alpha_{k_1}$ and $\beta_{k_1}$ are obtained as
\begin{equation}
\label{alphadot}
    \dot{\alpha}_{k_1}=-\frac{i\alpha_{k_1}^2}{2}+\frac{i\omega_{k_1}^2(t)}{2}
\end{equation}

\begin{equation}
    \label{betadot}
   i\dot{\beta}_{k_1}/\beta_{k_1}= \alpha_{k_1}/2  
\end{equation}
The expression for $\omega_{k_1}^2(t)$ is given by $\omega_{k_1}^2(t) = \lambda + (k_1 + qA_1(t))^2$, where $\lambda = m^2$, and, in \ref{alphadot} and \ref{betadot}, the dot represents the derivative with respect to time $`t'$. 

Next, we define and write 
\begin{equation}
    \label{alphaz}
    \alpha_{k_1}(t)=:\omega_{k_1}(t)\Bigg[\frac{1-z_{k_1}(t)}{1+z_{k_1}(t)}\Bigg]\quad 
\end{equation}
The canonical mode operators for the instantaneous frequency $\omega_{k_1}(t)$ are
\begin{eqnarray}
\label{op}
  \hat{a}_{k_1}(t) = \sqrt{\frac{\omega_{k_1}(t)}{2}}\phi_{k_1} + \frac{i}{\sqrt{2\omega_{k_1}(t)}}\Pi_{k_1} \\
   \hat{a}^\dagger_{k_1}(t) = \sqrt{\frac{\omega_{k_1}(t)}{2}}\phi^\dagger_{k_1} - \frac{i}{\sqrt{2\omega_{k_1}(t)}}\Pi^\dagger_{k_1}
\end{eqnarray}
Using these, the expectation value of the number operator in the Gaussian state \ref{psi} is
\begin{equation}
    \label{avgnk1}
    \langle n_{k_1}\rangle = \langle a^\dagger_k a_k\rangle=\frac{\langle|\Pi_{k_1}|^2\rangle}{2\omega_{k_1}}+\frac{\omega_{k_1}\langle|\phi_{k_1}|^2\rangle}{2}-\frac{1}{2}
\end{equation}
Computing the overlaps explicitly for the Gaussian wavefunction leads to a compact expression in terms of $z_{k_1}(t)$ \cite{Mahajan:2007qc, Mahajan:2007qg}:
%Initially, the system starts in the ground state with no particles present, but as time progresses, it departs from the instantaneous ground state. The particle content at any given time is determined by calculating the overlap between the evolving state and the instantaneous eigenstates that are adiabatically evolving at each moment. By computing this overlap following \cite{Mahajan:2007qc, Mahajan:2007qg}, one finds that the mean particle number per mode $k_1$ is
\begin{equation}
    \label{avgnk}
    \langle n_{k_1}  \rangle = \frac{|z_{k_1}(t)|^2}{1-|z_{k_1}(t)|^2}
\end{equation}
where $z_{k_1}(t)$ is defined in \ref{alphaz}. %Note for $\langle n_k\rangle$, we have simplified the subscript from $k_1$ to $k$.

%On computing the overlaps explicitly for the Gaussian wavefunction
%so that in terms of the variable $z_{k_1}(t)$, the evolution equation \ref{alphadot} becomes
%\begin{equation}
 %   \dot{z_{k_1}}+2i\omega_{k_1} z_{k_1}+\frac{\dot{\omega}_{k_1}}{\omega_{k_1}}(z_{k_1}^2-1)=0
%\end{equation}

The task now is to solve these dynamical equations with the appropriate initial conditions that define the ``vacuum" state at an initial time and then track the wavefunction's evolution. Since the variables are interdependent, solving for one allows us to deduce the others. To compute \ref{avgnk}, we solved \ref{alphadot} and then using the definition \ref{alphaz}, we find out $z_{k_1}(t)$ in terms of $\alpha_{k_1}(t)$ as
\begin{equation}
    \label{zdef}
    z_{k_1}(t) = \frac{\omega_{k_1}(t)-\alpha_{k_1}(t)}{\omega_{k_1}(t)+\alpha_{k_1}(t)}
\end{equation}

However, to solve \ref{alphadot}, we need to find the initial condition of $\alpha_{k_1}(t)$ at $t=0$. For this, we use the fact that initially the state is a ground state and has no particle content, which sets $|z_{k_1}(0)| =0$, and this sets 
\begin{equation}
    \label{initialalpha}
    \alpha_{k_1}(t=0) = \sqrt{\lambda +k_1^2} = \sqrt{m^2 + k_1^2}
\end{equation}

\begin{figure}
    \centering 
\includegraphics[scale=.55]{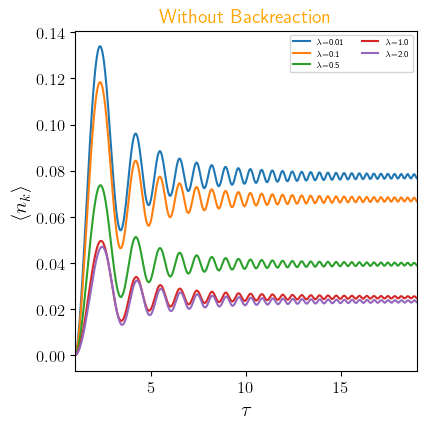}
\caption{ \small The plot shows $\langle n_k\rangle = \langle n_{-k}\rangle$ versus $\tau$ for different $\lambda$. Initially, $\langle n_k\rangle$ undergoes damped oscillations, settling to a finite value at large $\tau$. Smaller $\lambda$ yields stronger oscillations, while larger $\lambda$ reduces amplitude, accelerates saturation, and lowers the steady-state value. In $(1+1)-$dimensions, $\lambda = m^2$, and in $(1+3)-$dimensions, $\lambda = m^2 + |{\bf k}_p|^2$. The legend indicates the $\lambda$ values; the subscript $k_1$ is abbreviated to $k$.
}

\label{velocitywbr}
\end{figure}

Finally, after calculating $\alpha_{k_1}(t)$, we determined the particle number density $\langle n_{k_1} \rangle$ as given by equation \ref{avgnk}, using the definition in \ref{zdef}. The time evolution of this quantity is depicted in \ref{velocitywbr}, where $\langle n_{k_1} \rangle$ is plotted as a function of the dimensionless parameter $\tau = - \sqrt{q/E_0}A_1(t)$. Here $E_0$ represents the seed electric field with a constant strength set to $1$.

Using a similar approach, one can compute $\langle n_{-k_1} \rangle$ by replacing $\omega_{k_1}^2$ with $\omega_{-k_1}^2 = \lambda + (|k_1| - qA_1(t))^2$. The variation of $\langle n_{-k_1} \rangle$ with respect to the parameter $\tau$ is identical to that of $\langle n_{k_1} \rangle$. This suggests that in Minkowski spacetime, the number of particles with momentum $k_1$ is equal to the number of particles with momentum $-|k_1|$. %In this paper $k = |\mathbf{k}|$, where $\mathbf{k}$ represents the momentum vector of the particle.
%\subsection{$(1+3)-$ dimensional Minkowski spacetime}

In $(1+3)$-dimensional spacetime, using the same form of the gauge field $A_\mu$ that produces a homogeneous electric field along a single direction, the equations retain their structure, with the only change being that the parameter $\lambda$ is now given by $\lambda = |k_p|^2 + m^2$, where $k_p$ denotes the perpendicular momentum and $|k_p| = \sqrt{k_2^2 + k_3^2}$. In this case, the oscillation frequency that governs the time evolution of $\langle n_{k} \rangle$ (or $\langle n_{-k} \rangle$) becomes
\[
\omega_k^2 = |\mathbf{k}|^2 - 2q|k_1|A_1(t) + q^2 A_1^2(t).
\]
The evolution of the particle number expectation value follows the same form as given in \ref{avgnk}.
In both \((1+1)\)- and \((1+3)\)-dimensional Minkowski spacetimes, the particle number density \( \langle n_k(t) \rangle \) exhibits damped oscillations due to non-adiabatic evolution in a time-dependent electric field. The behavior is symmetric under \(|k| \leftrightarrow -|k|\), reflecting time-reversal invariance. In \((1+1)\) dimensions, particle production is efficient and saturates over time, while in \((1+3)\) dimensions, transverse momentum acts as an effective mass, suppressing production and increasing oscillation frequency.

Each momentum mode behaves like a harmonic oscillator with a time-dependent frequency. Non-adiabaticity causes deviations from the instantaneous vacuum, leading to squeezing and oscillations in \( \langle n_k(t) \rangle \). These variations arise from quantum interference, not the creation of real particles; however, the time-averaged particle number remains constant.

\begin{comment}
    
The physical momentum of a charged particle of momentum $k$ with respect to the co-moving observer is given by
\begin{equation}
    p^\pm=(k\pm q A_1(t))
\end{equation}
Therefore, the drift velocity of particles and antiparticles created by the background electric field is
\begin{equation}
    \label{vde}
    v= \frac{p^\pm}{\sqrt{m^2+p^{\pm \,2}}}=\frac{p^\pm}{\sqrt{\lambda+p^{\pm \,2}}}
\end{equation}

The variation of particle velocity \ref{vde} concerning the parameter $\tau$ is shown in \ref{velocitywbr}, it increases monotonically with the increase in parameter $\tau$.
\end{comment}

\section{Influence of Backreaction on Pair Creation Processes}
\label{Schwinger effect with backreaction}
%\subsection{$(1+1)-$ dimensional de Sitter spacetime}
Having analysed Schwinger pair production without backreaction in the previous section, we now incorporate the feedback of the created particles on the background field through a self–consistent semiclassical evolution. Our primary objective is to determine how this backreaction modifies the electric field and the associated observables. We start from the semiclassical evolution equation:
\begin{equation}
    \label{EFMeq}
    -\frac{dE(t)}{dt}= \langle \hat{J}^\mu_Q \rangle
\end{equation}
where $\hat{J}^\mu_Q$ is the current operator of the complex scalar field and its expectation value is taken in evolving the quantum state of the field dictated by the dynamical background electric field. It is defined as
\begin{equation}
\begin{split}
\label{current}
    \hat{J}^\mu_Q = \eta^{\mu\nu}[-iq(\hat{\phi}^\dagger(\partial_\nu \hat{\phi})-(\partial_\mu \hat{\phi}^\dagger)\hat{\phi}) -2 q^2A_\nu (\hat{\phi}^\dagger \hat{\phi}) ]
    \end{split}
\end{equation}
Note that the $\mu =0$ component of the current is
zero, meaning that no net charge is created. We computed the non-zero spatial component of current, which is given as
\begin{equation}
    \label{current0}
     \hat{J}^1_Q=iq(\hat{\phi}^\dagger(\partial_1 \hat{\phi})-(\partial_1 \hat{\phi}^\dagger)\hat{\phi}) +2 q^2A_1(t) (\hat{\phi}^\dagger \hat{\phi})
\end{equation}
In the vacuum state, it is given as

\begin{equation}
    \label{current1}
    \langle \hat{J}^1_Q\rangle = q\int^{\infty}
_{-\infty} \frac{dk_1}{2\pi} k_1 \langle|\phi_{k_1}|^2\rangle +2 q^2 A_1(t)\int^{\infty}_{-\infty}\frac{dk_1}{2\pi}\langle|\phi_{k_1}|^2\rangle=4q^2A_1(t)\int^{\infty}_{0}\frac{dk_1}{2\pi}\langle|\phi_{k_1}|^2\rangle
\end{equation}

%\begin{equation}
%\label{currentds1}
 %   \langle \hat{J}^1_{Q}\rangle = 4 q^2 A_1(t )\int_{0}^{\infty} \frac{dk_1}{2\pi} \Big(\frac{1}{Re(\alpha_{k_1,dS}(\eta))}+\frac{qm^2}{2(k_1^2+m^2)^{\frac{3}{2}}} A_1(t)\Big)
%\end{equation}

where we used the fact that $\phi^\dagger_{k_1}\phi_{k_1}=\phi^\dagger_{-k_1}\phi_{-k_1} /; (|\phi_{k_1}|^2 = |\phi_{-k_1}|^2)$. %

On substituting \ref{current1} in \ref{EFMeq}, the equation of motion of electric field is given as 
\begin{equation}
    \label{EFeqm}
    \frac{d E}{dt}=-4q^2A_1(t)\int^{\infty}_{0}\frac{dk_1}{2\pi}\langle|\phi_{k_1}|^2\rangle
\end{equation}
This equation can be obtained by evaluating the Hamilton's equation of motion for the electric field with respect to the effective Hamiltonian in \ref{Hamiltoniandensity}. The next step involves computing the expectation value of $|\phi_{k_1}|^2$ using the Gaussian ansatz for the wavefunction defined in \ref{psi}, which is given by:
\begin{equation}
    \label{expphi}
    \langle |\phi_{k_1}|^2 \rangle = \frac{1}{4\text{Re}(\alpha_{k_1})}
\end{equation}
On substituting \ref{expphi} in \ref{EFeqm}, we obtain

\begin{equation}
    \label{EFavg}
    \Dot{E}=  \frac{d E}{dt}=-q^2 A_1(t)\int_0^\infty \frac{dk_1}{2\pi}\frac{1}{\text{Re}(\alpha_{k_1} (t))}
\end{equation}

The integral in \ref{EFavg} exhibits ultraviolet divergence. To address this, we performed the integration by assuming a one-dimensional lattice with lattice length $\ell$, which results in
\begin{equation}
    \label{EFavglattice}
   \Dot{E}=-\frac{q^2 A_1(t)}{\ell}\sum_{n}\frac{1}{\text{Re}(\alpha_{k_{1_n}}(t))}
\end{equation}
where the summation is over all lattice points denoted by $n$. To account for possible ultraviolet (UV) divergences in our computations, we have implemented a discretization scheme. In particular, a finite lattice spacing is introduced, which restricts the highest accessible momentum modes and serves as a natural UV cutoff. As a result, the numerical evolution is well-defined and quantities like the induced current and particle number density are finite. In addition, we have confirmed that our findings are robust across both schemes and that the discretization has no effect on the physical outcome when employing the regularized action suggested in earlier studies~\cite{BRSE6, vilenkin}.

In contrast to mode-based methods that require adiabatic regularization to subtract divergent vacuum contributions as shown in \cite{BRSE1, BRSE2, BRSE3, BRSE4, BRSE5, BRSE6, BRSE7, BRSE8}, our semiclassical framework evolves a finite quantum state via the time-dependent Schrödinger equation. This preserves the entire non-adiabatic dynamics of the system, including tunnelling effects, plasma oscillations, and the influence of cosmological expansion, and enables us to compute observables directly from expectation values without the need for subtractive procedures.

\begin{figure}
    \centering
     \includegraphics[scale=.55]{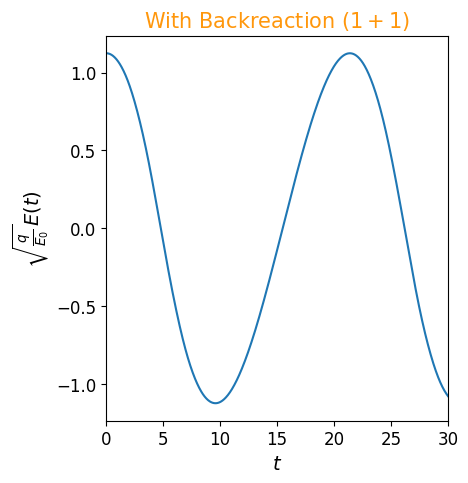}\hspace{0.7cm}
     \includegraphics[scale=.55]{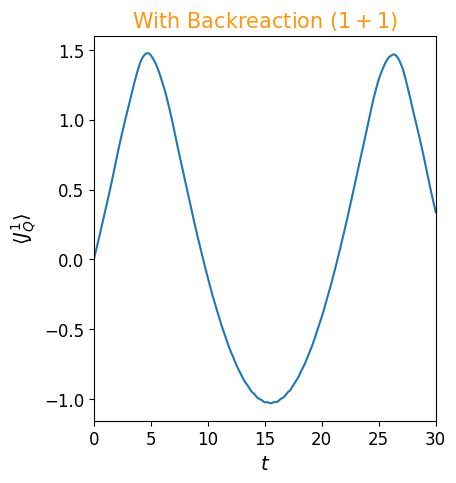}\\\vspace{0.5cm}
     \includegraphics[scale=.55]{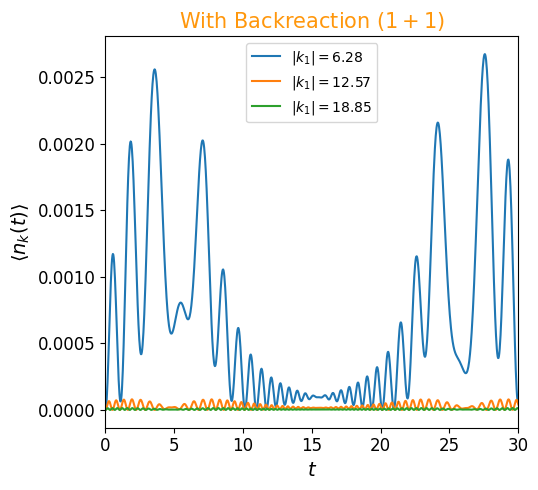}\hspace{0.6cm}
     \includegraphics[scale=.55]{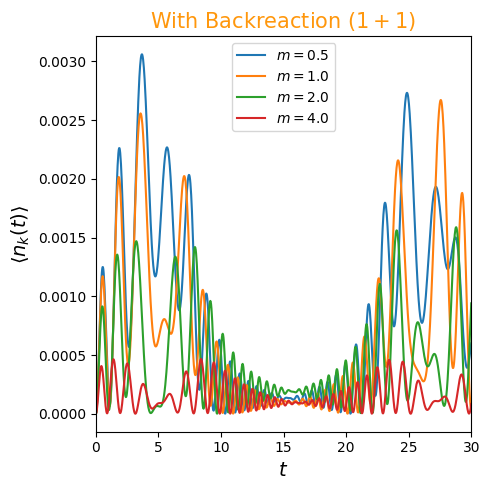}
    \caption{ \small The first row shows (left) the time evolution of the electric field and (right) the corresponding current 
$\langle J^1_Q \rangle$ generated by particle-antiparticle pairs in $(1+1)-$ dimensions. 
The second row shows the time evolution of the particle (or antiparticle) number: (left) for various $|k_{1}|$ 
at fixed $E_{0}=4.0$, $q=0.316$, $m=1.0$; and (right) for various $m$ at fixed $E_{0}=4.0$, 
$q=0.316$, $|k_{1}|=1$. (Here the subscript $k_{1}$ in $\langle n_{k_{1}}\rangle$ is abbreviated to $k$.)
}
    \label{fig:BR}
\end{figure}

On defining $\tau=-\sqrt{q/E_0}A_1(t)$, one can rewrite above equation as
\begin{equation}
    \label{ddtau}
    \Ddot{\tau}=-\frac{q^{2} \tau}{\ell}\sum_{n}\frac{1}{\text{Re}(\alpha_{k_{1_n}}(t))}
\end{equation}
Note that the electric field is defined as
\begin{equation}
    \label{tauEF}
    E(t)=\sqrt{\frac{E_0}{q}}\dot{\tau}
\end{equation}
and in terms of $\tau$ \ref{alphadot} becomes
\begin{equation}
    \label{alphadotBR}
    \Dot{\alpha_{k_{1}}}=-\frac{i\alpha_{k_1}^2}{2}+\frac{im^2}{2}+\frac{i (k_1-\sqrt{q E_0} \tau)^2}{2}
\end{equation}
The equations \ref{ddtau}, \ref{tauEF}, and \ref{alphadotBR} form a coupled system giving self-consistent dynamics. Solving these equations enables us to determine the evolution of the electric field, the current density $\langle J^1_Q \rangle$, and the average particle number density with respect to the time parameter $t$, as shown in \ref{fig:BR}. When the backreaction from the created particles is taken into account, the plasma oscillations emerge in the electric field. These oscillations in the electric field and current density result in specific modes experiencing multiple particle creation events and, at times, particle annihilation as observed in \cite{BRSE1, BRSE6}. In contrast to adiabatic methods like \cite{BRSE1, BRSE6}, which forecast a transient plateau in $\langle J^1_Q \rangle$, our non-perturbative approach generates plasma oscillations instantly.  This difference arises from (i) the strong, self-consistent coupling between field dynamics and particle generation from the beginning, and (ii) the use of instantaneous eigenstates that suppress quasi-steady behavior and capture rapid field fluctuations.  Our framework uses the time-dependent Schrödinger equation to track the full real-time evolution of the quantum state, allowing for immediate feedback from created particles on the electric field, in contrast to previous studies that rely on slowly varying backgrounds and allow the system to momentarily settle into a quasi-steady state.  This continuous, nonlinear feedback results in extended oscillating behaviour rather than the formation of a plateau. This constant, nonlinear feedback prevents the creation of a plateau, instead resulting in prolonged oscillating behaviour. These findings highlight the importance of backreaction modeling choices, emphasizing the need for non-perturbative, fully dynamical treatments to adequately capture the rich and strongly coupled dynamics of pair creation in intense fields. These Minkowski–space results set the stage for our subsequent analysis of backreaction in de Sitter spacetime, where expansion further modifies the plasma oscillations.

\begin{figure}
    \centering
        \includegraphics[scale=.52]{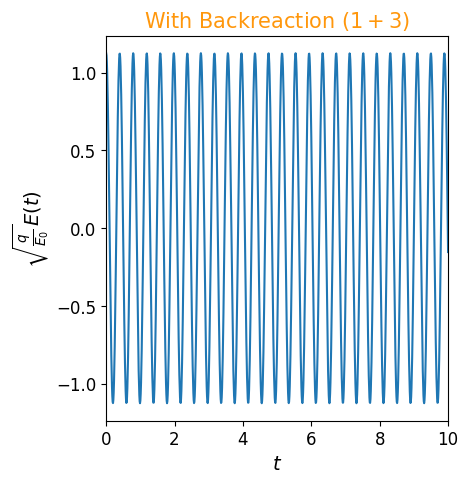}\hspace{0.10cm}
     \includegraphics[scale=.52]{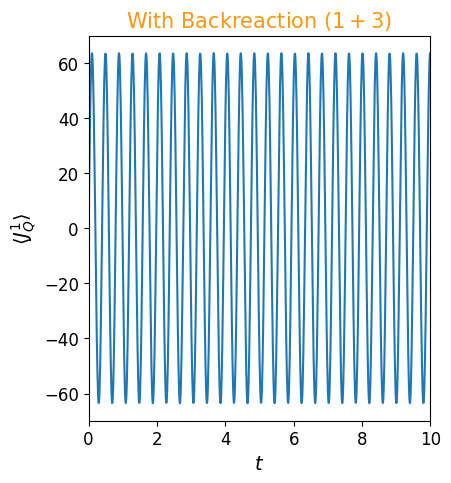} \\
      \includegraphics[scale=.48]{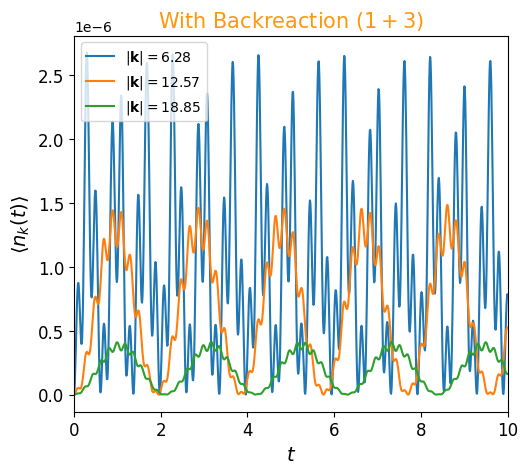}\hspace{0.10cm}
     \includegraphics[scale=.48]{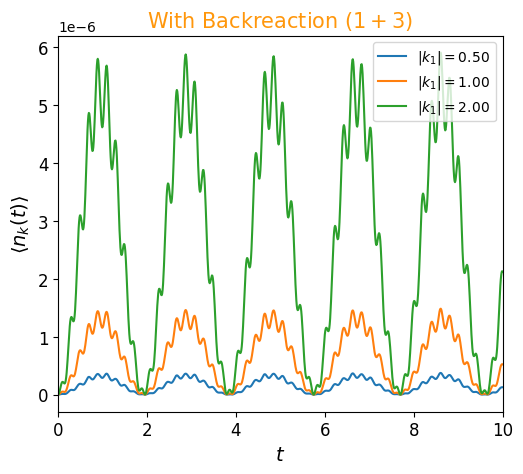}\\\includegraphics[scale=.48]{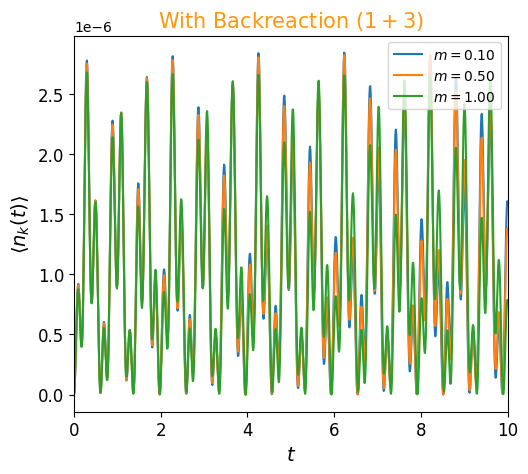}
   \caption{ \small The figure shows particle-antiparticle dynamics in $(1+3)-$dimensions Minkowski spacetime.  
First row: time evolution of the electric field (left) and induced current $\langle J^1_Q \rangle$ (right) for $E_0 = 4.0$, $|k_1| = 1$, $q = 0.316$, and $m = 1.0$.  
Second row: evolution of particle number $\langle n_k \rangle$; left: varying $|\mathbf{k}|$ with $E_0 = 4.0$, $q = 0.316$, $|k_1| = 1$, $m = 1.0$; right: varying $|k_1|$ with $E_0 = 4.0$, $q = 0.316$, $|\mathbf{k}| = 1$, $m = 1.0$.  
Last row: $\langle n_k \rangle$ for different $|k_1|$ under the same settings as the second-row right plot.}
    \label{fig:BR1}
\end{figure}
%\subsection{$(1+3)$-dimensional Minkowski spacetime}
For $(1+3)-$dimensional spacetime, the non-zero spatial component of current in the vacuum state is given as
\begin{equation}
    \langle \hat{J}^1_Q \rangle = 2 q^2 A_1(t) \int \frac{d^3k}{(2\pi)^3} \langle |\phi_k|^2\rangle
\end{equation}
Using spherical coordinates, the integral simplifies to  
\begin{equation}
    \langle \hat{J}^1_Q \rangle = \frac{q^2 A_1(t)}{\pi^2} \int_0^\infty  d k  \;k^2\, \langle |\phi_{k}|^2 \rangle
\end{equation}

here $k$ is the modulus of three-momentum $\textbf{k}$. For a system of finite size \( \ell \), the allowed momenta are discretized as  
\begin{equation}
    k_{n} = \frac{2\pi n}{\ell}, \quad n \in \mathbb{Z}^+.
\end{equation}
Replacing the integral with a sum, we obtain  
\begin{equation}
    \langle \hat{J}^1_Q \rangle = \frac{q^2 A_1(t)}{\pi \ell} \sum_n k_n^2 \langle |\phi_{k_{n}}|^2 \rangle.
\end{equation}

Using \ref{expphi}, the final expression:  
\begin{equation}
    \langle \hat{J}^1_Q \rangle = \frac{q^2 A_1(t)}{4\pi \ell} \sum_n \frac{k_n^2}{\text{Re}(\alpha_{k_{n}})}.
\end{equation}

The time evolution of the electric field, the current density \(\langle \hat{J}^1_Q \rangle\), and the average particle number density in the \((1+3)\)-dimensional Minkowski spacetime are illustrates in \ref{fig:BR1}. In Minkowski spacetime, the inclusion of backreaction leads to sustained plasma-like oscillations in the electric field, driven by the self-consistent feedback between the produced particle–antiparticle pairs and the field itself. The number density \( \langle n_k \rangle \) in both \((1+1)\)- and \((1+3)\)-dimensional settings exhibits such oscillations following an initial growth phase. In \((1+1)\) dimensions, these oscillations occur around a nearly constant amplitude, reflecting a dynamic balance between pair production and field regeneration. In contrast, the \((1+3)\)-dimensional case features higher oscillation frequencies due to the enhanced contribution from transverse momentum modes, which increase the effective frequency of each mode and lead to a denser phase space. This results in stronger current–field feedback and faster plasma oscillations, although the amplitude is slightly reduced owing to the larger effective mass. In both cases, the dynamics remain symmetric under \(k \leftrightarrow -k\), and no net damping is observed over the time scales considered. These results highlight the crucial role of dimensionality and momentum structure in shaping the real-time evolution of backreacting quantum fields.

\section{Schwinger effect in the cosmological de Sitter spacetime}
\label{Schwinger effect in the cosmological de Sitter spacetime}
%\subsection{$(1+1)$- dimensional de Sitter spacetime}
Building on the analysis of backreaction in Minkowski spacetime, we now extend our framework to a complex scalar field in a background electric field on $(1+1)$ conformally flat de Sitter spacetime given as 

\begin{equation}
    \label{1+1Ds}
    ds^2=a^2(\eta)(d\eta^2-dx^2)
\end{equation}
where $a(\eta)$ is the scale parameter and $\eta$ is  conformal time. For de Sitter space, one has $\eta = -1/Ha$ with constant Hubble parameter $H$. We revisit Schwinger pair production in this background, analysing both the case without and with backreaction using the canonical approach.

The Hamiltonian density of a complex scalar field $\phi$ of mass $m$ coupled to an external gauge field $A_\mu(x)= (0,A_1(\eta))$ is 
\begin{equation}
    \label{dShamiltonian}
    \mathcal{H_{\text{dS}}}=\frac{E^2(\eta)}{2}+\frac{1}{2}\Big[\Pi^\dagger\Pi+(\partial_1- qA_1(\eta))\phi^\dagger(\partial_1+iqA_1(\eta))\phi+m^2a^2(\eta)\phi^\dagger\phi\Big]
\end{equation}

where $\Pi$ and $E$ are the conjugate moments corresponding to $\phi$ and $A_1$, respectively. We choose $A_\mu(x)= (0,A_1(\eta))$ which leads to a non-zero electric field given by

\begin{eqnarray}
    \label{aprime}
    F_{01}&=& -\frac{\partial A_1}{\partial \eta} = \sqrt{-g}E(\eta)=E(\eta)a^2(\eta)\\
    E(\eta)&=&-\frac{1}{a^2(\eta)}\frac{\partial A_1}{\partial \eta}
\end{eqnarray}

where $g$ is the determinant of the metric \ref{1+1Ds} \cite{vilenkin}. In this scenario also, we consider an initial seed electric field $E=E_0$ at $\eta=-\eta_0$, where $\eta_0>0$, which imposes the condition
\begin{equation}
    \label{EintialdS}
    E_0=-\left. \frac{1}{a^2(\eta)}\frac{\partial A_1(\eta)}{\partial \eta} \right|_{\eta=-\eta_0}
\end{equation}

The total Hamiltonian is 
\begin{equation}
    \label{dSHT}
    H_{dS}= \int dx \Big[\frac{E^2(\eta)}{2}+\frac{1}{2}\Big(\Pi^\dagger\Pi+(\partial_1-iqA_1(\eta))\phi^\dagger(\partial_1+iqA_1(\eta))\phi+m^2a^2(\eta)\phi^\dagger\phi\Big)\Big]=H_{1,dS}+H_{2,dS}
\end{equation}
where the separable components are defined as follows: 
\begin{eqnarray}
\label{CQdS}
    H_{1,dS}&=&\frac{1}{2}\int dx\;E^2(\eta)\\
    H_{2,dS}&=& \frac{1}{2}\int dx  \Big(\Pi^\dagger\Pi+(\partial_1-iqA_1(\eta))\phi^\dagger(\partial_1+iqA_1(\eta))\phi+m^2a^2(\eta)\phi^\dagger\phi\Big)
\end{eqnarray}
Here, $H_{1,dS}$ is the purely classical part, while $H_{2,dS}$ contains the quantum field in which the electric field enters as a $c$-number.

In the Fourier space,
\begin{equation}
    \label{HkdS}
    H_{2,dS}=\frac{1}{2}\int \frac{dk_1}{2\pi}\Big[\Pi^\dagger_{k_1}\Pi_{k_1}+(k_1+qA_1(\eta))^2\phi^\dagger_{k_1}\phi_{k_1}+m^2a^2(\eta)\phi^\dagger_{k_1}\phi_{k_1}\Big]=H_{2,dS}(k_1) \oplus H_{2,dS}(-k_1)
\end{equation}
here 
\begin{eqnarray}
    H_{2,dS}(k_1) &=& \frac{1}{2}\int_{0}^{\infty} \frac{dk_1}{2\pi}\Big[|\Pi_{k_1}|^2+(k_1+qA_1(\eta))^2|\phi_{k_1}|^2+m^2a^2(\eta)|\phi_{k_1}|^2\Big]\\
       H_{2,dS}(-k_1) &=& \frac{1}{2}\int_{0}^{\infty} \frac{dk_1}{2\pi}\Big[ |\Pi_{-k_1}|^2+(|k_1|-qA_1(\eta))^2|\phi_{-k_1}|^2+m^2a^2(\eta)|\phi_{-k_1}|^2\Big] 
\end{eqnarray}
We take the product state 
\begin{equation}
    \label{psidefdS}
    \psi_{dS} = \prod_{k_1,-k_1} ( \psi_{k_1,dS} \otimes \psi_{-k_1,dS} )
\end{equation}
same as \ref{psidef} and, because modes do not mix, work with a single.

The time-dependent Schrodinger equation (TDSE) is
\begin{equation}
    \label{tdse1}
    i \frac{\partial}{\partial t} (\psi_{k_1,dS} \otimes \psi_{-k_1,dS})= (\hat{h}_{k_1,dS}\otimes \hat{I}_{-k_1,dS} \oplus \hat{I}_{k_1,dS}\otimes \hat{h}_{-k_1,dS} )(\psi_{k_1,dS} \otimes \psi_{-k_1,dS})
\end{equation}
here $\hat{h}_{k_1,dS}$ and $\hat{h}_{k_1,dS}$ are given as
\begin{eqnarray}
 \label{hathk}
    \hat{h}_{k_1,dS}=\frac{1}{2}\Big(|\Pi_{k_1}|^2+(k_1+qA_1(\eta))^2|\phi_{k_1}|^2+m^2a^2(\eta)|\phi_{k_1}|^2\Big)\\
    \label{hathmk}
    \hat{h}_{-k_1,dS}=\frac{1}{2}\Big(|\Pi_{-k_1}|^2+(|k_1|-qA_1(\eta))^2|\phi_{-k_1}|^2+m^2a^2(\eta)|\phi_{-k_1}|^2\Big)
\end{eqnarray}

On decoupling the time-dependent Schrodinger equation for $\psi_{k,dS}$ is
\begin{equation}
    \label{TDSEdS}
    i\frac{\partial}{\partial\eta}\psi_{k_1,dS}(\phi_{k_1},A_1(\eta),\eta)=\hat{h}_{k_1,dS}(\eta)\psi_{k_1,dS}(\phi_{k_1},A_1(\eta),\eta)
\end{equation}
We use the Gaussian ansatz of wavefunction $\psi$ given as
\begin{equation}
    \label{psidS}
    \psi(\phi_{k_1},A_1(\eta),\eta)=\beta_{k_1,dS}(\eta)e^{-\alpha_{k_1,dS}{(\eta)}\phi_{k_1}\phi_{k_1}^\dagger}
\end{equation}
On substituting \ref{psidS} in \ref{TDSEdS}, equations of motion for $\alpha_{k,dS}$ and $\beta_{k,dS}$ are
\begin{equation}
    \label{alphadedS}
    \alpha^\prime_{k_1,dS}=-\frac{i\alpha^2_{k_1,dS}}{2}+\frac{i\omega_{k_1,dS}^2(\eta)}{2}
\end{equation}
\begin{equation}
    \label{betadS}
    i\beta_{k_1,dS}^\prime/\beta_{k_1,dS}=\alpha_{k_1,dS}/2
\end{equation}
where $\omega_{k_1,dS}^2(\eta)=(k_1+q A_1(\eta))^2+m^2a^2$, we redefine it as
\begin{equation}
    \label{nu}
    \omega_{k_1,dS}^2(\eta)=|k_1|^2[(1+L \xi)^2+M^2 \xi^2]
\end{equation}
where \( L \) and \( M \) are defined as  
\begin{equation}
L = \frac{qE}{H^2}, \quad M = \frac{m}{H},
\end{equation}  
which can be referred to as the rescaled electric field and mass, respectively. Additionally, we define  
\begin{equation}
\xi = \frac{Ha(\eta)}{|k_1|}.
\end{equation}  
Its derivative with respect to the conformal time \( \eta \) is given by  
\begin{equation}
\label{xieta}
\xi' = |k_1| \xi^2.
\end{equation}
Following the same parametrization as in the Minkowski case, we define
\begin{equation}
\label{zeqdS}
    \alpha_{k_1,dS}(\eta)=\omega_{k_1,dS}\Bigg(\frac{1-z_{k_1,dS}(\eta)}{1+z_{k_1,dS}(\eta)}\Bigg)
\end{equation}
%On substituting it in \ref{alphadedS}, we obtain
%\begin{equation}
%\label{zdedS}
 %   z^\prime_{k_1,dS}+2i\omega_{k_1} z_{k_1,dS}+\frac{\dot{\omega}_{k_1,dS}}{\omega_{k_1,dS}}(z_{k_1,dS}^2-1)=0
%\end{equation}
Likewise \ref{avgnk}, the average number of particles with momentum $k$ is
\begin{equation}
    \label{avgnkdS}
    \langle n_{k,dS} \rangle = \frac{|z_{k_1,dS}|^2}{1-|z_{k_1,dS}|^2}
\end{equation}
For computing \ref{avgnkdS}, we solved \ref{alphadedS} and then using the definition \ref{zeqdS}, we find out $z_{k_1,dS}(\eta)$ in terms of $\alpha_{k_1,dS}(\eta)$ as
\begin{equation}
    \label{z}
    z_{k_1,dS}=\frac{\omega_{k_1,dS}-\alpha_{k_1,dS}}{\omega_{k_1,dS}+\alpha_{k_1,dS}}
\end{equation}

\begin{figure}
    \centering
    \includegraphics[scale=.40]{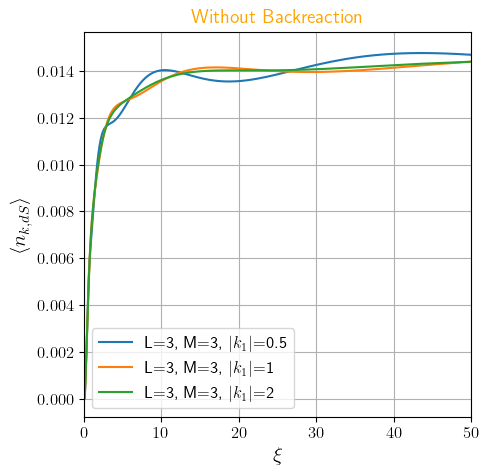}\hspace{0.1cm}
\includegraphics[scale=.40]{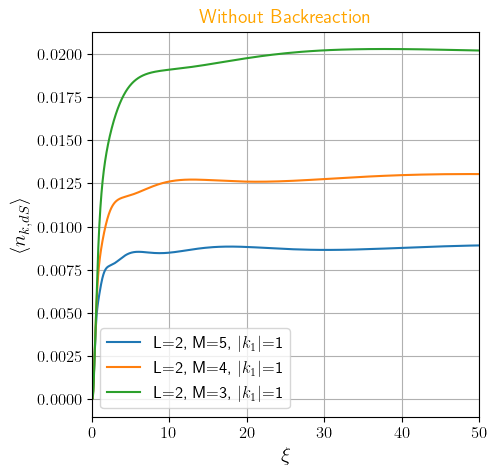} \hspace{0.1cm}
\includegraphics[scale=.40]{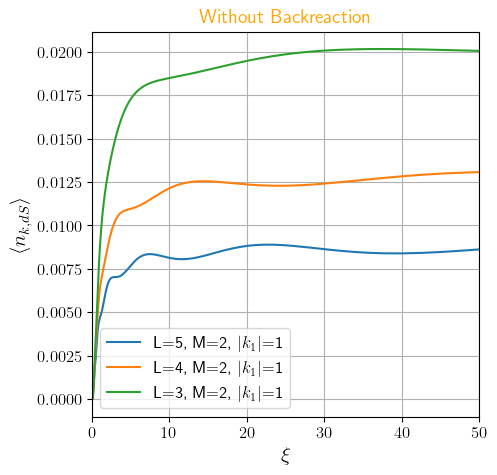} \\ \vspace{0.1cm}
\includegraphics[scale=.40]{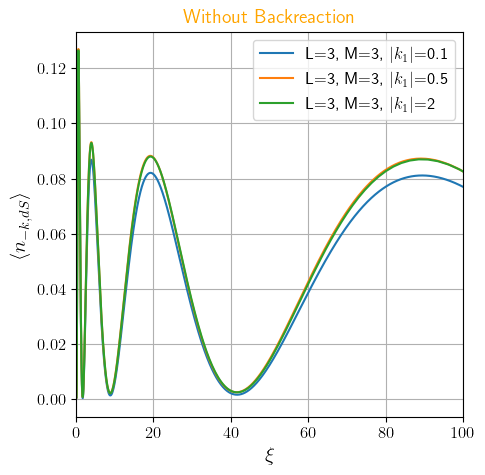}\hspace{0.1cm}
\includegraphics[scale=.40]{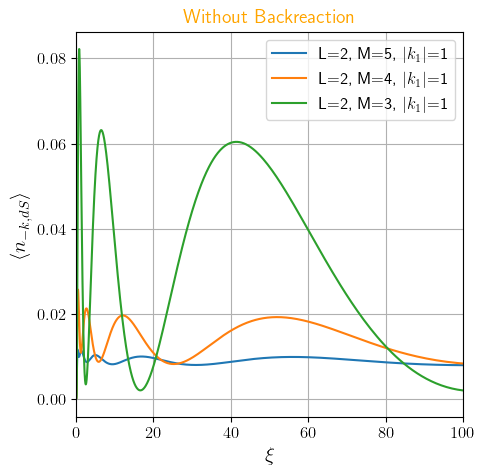} \hspace{0.1cm}
\includegraphics[scale=.40]{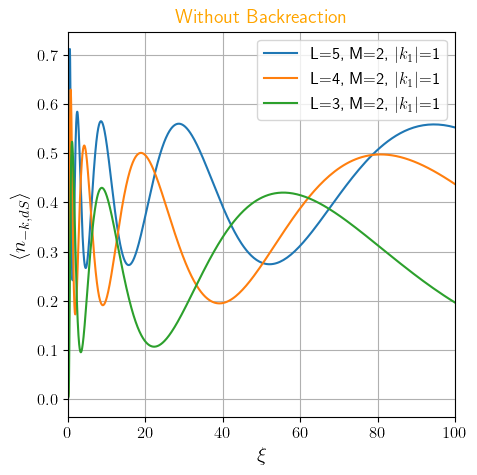} \\\vspace{0.1cm}
\includegraphics[scale=.30]{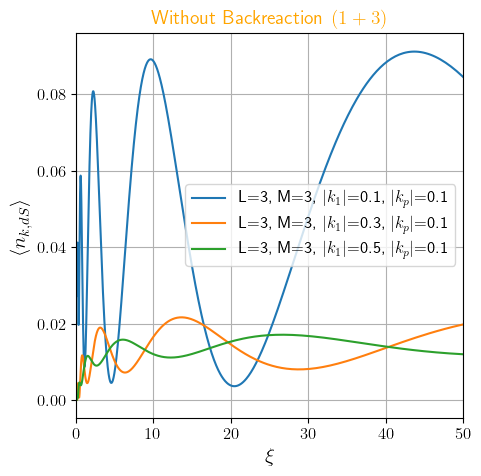}\hspace{0.1cm}
\includegraphics[scale=.30]{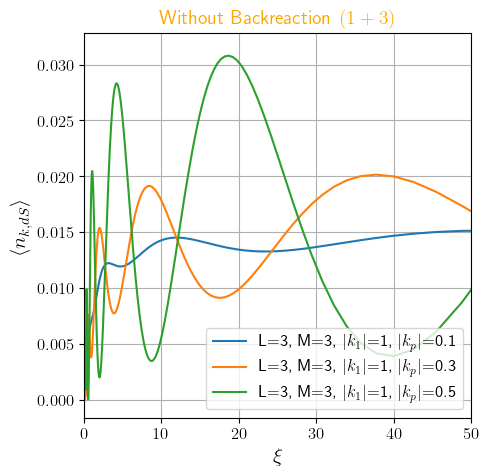}\hspace{0.1cm}
\includegraphics[scale=.30]{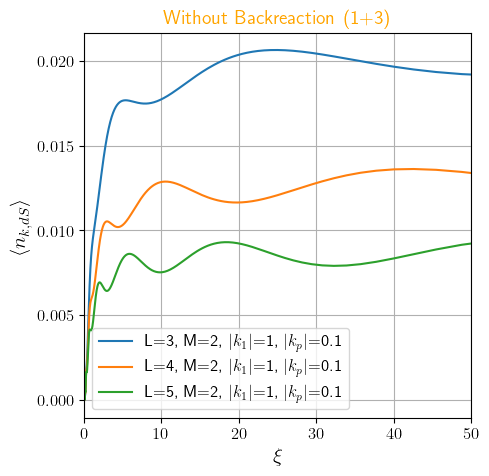}\hspace{0.1cm}
\includegraphics[scale=.30]{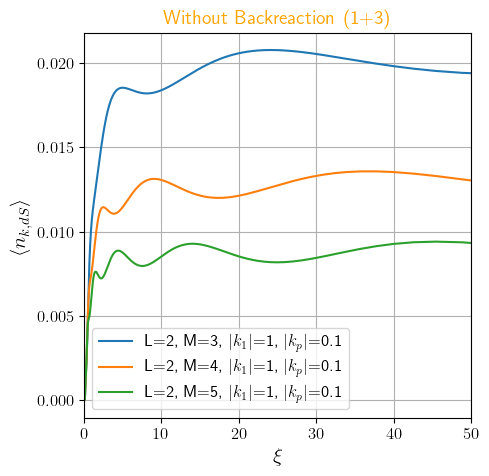}\\\vspace{0.1cm}
\includegraphics[scale=.30]{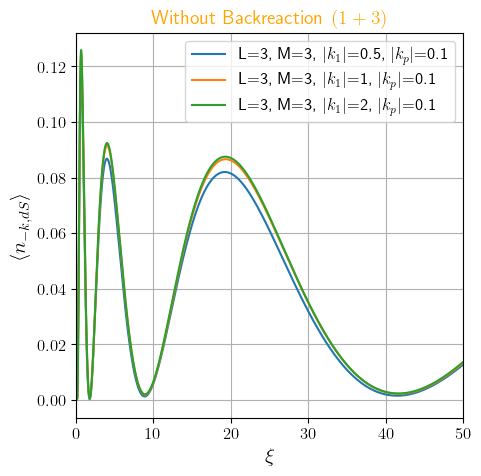}\hspace{0.1cm}
\includegraphics[scale=.30]{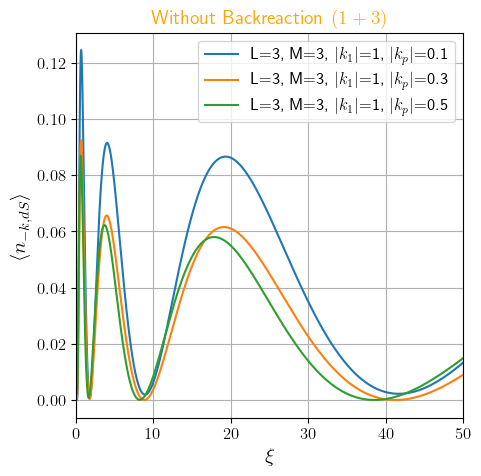}\hspace{0.1cm}
\includegraphics[scale=.30]{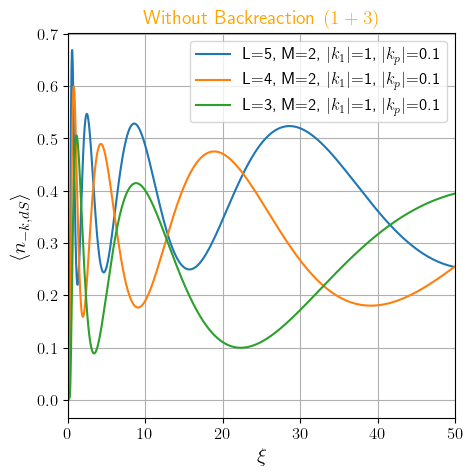}\hspace{0.1cm}
\includegraphics[scale=.30]{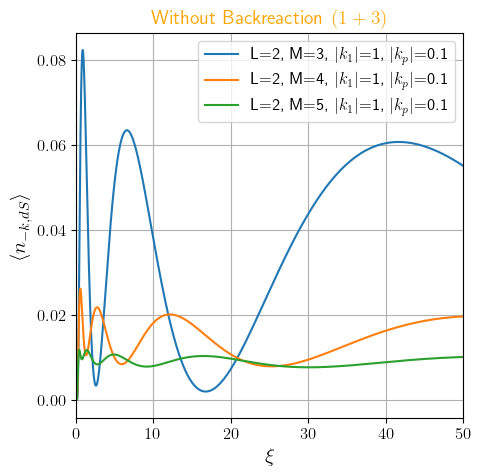}
\caption{ \small The plots show $\langle n_{k, \text{dS}} \rangle$ (rows 1 and 3) and $\langle n_{-k, \text{dS}} \rangle$ (rows 2 and 4) versus $\xi$. In $(1+1)-$dimensions (rows 1--2), the left, middle, and right plots correspond to varying $|k_1|$, $M$, and $L$, respectively. In $(1+3)-$dimensions (rows 3--4), the left plots show $|k_1|$ dependence, the middle plots vary $|\mathbf{k_p}|$ and $L$, and the right plots vary $M$.
 }
    \label{DS1}
\end{figure}

We impose the initial vacuum following \cite{Sharma:2017ivh}.
In the asymptotic past  \( \eta \to -\infty \) (or equivalently \( a \to 0 \)), the frequency reduces to \( \omega_{k, \text{dS}} = |k_1| \), so we choose the Bunch–Davies state with \(\alpha_{k_1, \text{dS}}(\eta \to -\infty) = |k_1| \). Unlike Minkowski space, adiabaticity at late times is not automatic; we restrict to \( L^2 + M^2 \gg 1 \) to ensure an adiabatic regime.

For $-k_1$ the frequency becomes $\omega_{-k_1,dS}^2 = |k_1|^2[(1-L\xi)^2+M^2\xi^2] $. Consequently, $\langle n_{-k,dS}\rangle$ differs from $\langle n_{k,dS}\rangle$, reflecting the loss of $k \leftrightarrow -k$ symmetry in the de Sitter space; for more details, we refer our reader to \cite{vilenkin}. The variation of the number density of particles with momentum $k_1$ and $-|k_1|$ concerning parameter $\xi$ has been plotted in \ref{DS1}, respectively.

%We adopt the initial condition for the vacuum state based on the formulation presented in \cite{Sharma:2017ivh}. In the asymptotic past, as \( \eta \to -\infty \) (or equivalently \( a \to 0 \)), we find that \( \omega_{k, \text{dS}} = |k_1| \), allowing us to define a vacuum state in this limit. This vacuum state is equivalent to the Bunch-Davies vacuum. Consequently, the initial condition for \( \alpha_{k, \text{dS}}(\eta) \) is set by \( \alpha_{k_1, \text{dS}}(\eta \to -\infty) = |k_1| \). 

%Unlike in Minkowski spacetime, where adiabatic behaviour is typically guaranteed, the adiabatic regime at late times is not assured in this case. To define the late-time adiabatic regime, we impose the condition \( L^2 + M^2 \gg 1 \); otherwise, the evolution becomes non-adiabatic.

%Following a similar procedure, one can compute $\langle n_{-k,dS}\rangle$ where $\omega_{-k_1,dS}^2 = |k_1|^2[(1-L\xi)^2+M^2\xi^2] $. The variation of $\langle n_{-k,dS}\rangle$ with respect to the parameter $\xi$ is different compared to that of $\langle n_{k, dS} \rangle$.  This indicates that in the de Sitter spacetime, the number of particles with momentum $k_1$ is not equal to the number of particles with momentum $-|k_1|$, for more details, we refer our reader to \cite{vilenkin}. The variation of the number density of particles with momentum $k_1$ and $-|k_1|$ concerning parameter $\xi$ has been plotted in \ref{DS1}, respectively.
%\subsection{$(1+3)-$ dimensional de Sitter spacetime}

In $(1+3)$-dimensional spacetime, using the same form of the gauge field $A_\mu$ that generates a homogeneous electric field in a single direction, the structure of the equations remains unchanged. However, the frequency term $\omega_{k,\text{dS}}^2(\eta)$ is now given by
\[
\omega_{k,\text{dS}}^2(\eta) = |k_1|^2\left[(1 + L\xi)^2 + M^2\xi^2\right] + |{\bf k_p}|^2 = |{\bf k}|^2 + |k_1|^2\left[(1 + L\xi)^2 + M^2\xi^2 - 1\right],
\]
with ${\bf k_p}$ the momentum transverse to the field. The particle number still follows \ref{avgnkdS}.

As in the Minkowski analysis, each Fourier mode in de Sitter behaves as a time-dependent harmonic oscillator whose non-adiabatic evolution drives it into a squeezed state. Consequently, in both $(1+1)$- and $(1+3)$-dimensional de Sitter spacetimes the particle number $\langle n_{k}\rangle$ initially rises rapidly and then saturates, signalling a finite production epoch, with residual oscillations arising from the same interference mechanism described for Minkowski space. The key difference is that, unlike Minkowski space where $k\!\leftrightarrow\!-k$ symmetry holds and plasma-like oscillations appear mainly through backreaction, the de Sitter background itself breaks this symmetry, giving $\langle n_{k}\rangle\neq\langle n_{-k}\rangle$ even before backreaction is included. In fact, because the effective frequencies $\omega_{k}(\eta)$ for positive and negative momenta differ due to the combined electric field and expansion, the two signs of $k$ undergo distinct non-adiabatic mixing and phase evolution. This asymmetry reflects the fact that the electric field and cosmological expansion create a time-dependent potential with different forward and backward tunnelling probabilities for charged modes. By redshifting the momenta and modifying the effective barrier height, curvature alone changes the tunnelling dynamics and thus produces distinct particle-number evolutions for positive and negative momenta~\cite{vilenkin}.

%In both $(1+1)-$ and $(1+3)-$dimensional de Sitter spacetimes, the particle number density $\langle n_k\rangle$ typically exhibits rapid growth at early times, followed by saturation, indicating a finite period of particle production. Oscillatory features that appear in certain parameter regimes arise from quantum interference and non-adiabatic evolution in the time-dependent background. In particular, differences in the behavior of $\langle n_k\rangle$ and $\langle n_{-k}\rangle$ can be attributed to the distinction between forward tunnelling (downward) and backward tunnelling (upward) in the presence of an electric field on an expanding background, as observed earlier in \cite{vilenkin}. The de Sitter expansion influences the effective potential experienced by different momentum modes, modifying their tunnelling probabilities and leading to observable differences in particle number evolution. In contrast, such effects are absent in Minkowski spacetime, where symmetry between $k$ and $-k$ is preserved, and plasma-like oscillations emerge predominantly through backreaction. These findings demonstrate that spacetime curvature modifies the tunnelling dynamics compared to the flat case and produces distinct particle number evolution for positive and negative momenta, even before including backreaction effects.

\section{Backreaction Dynamics in de Sitter Spacetime}
\label{Schwinger effect with backreaction in de Sitter spacetime}
%\subsection{$(1+1)-$ dimensional de Sitter spacetime}
We now include the backreaction of the created particles on both the mean number density and the electric field by deriving the electric-field equation of motion from \ref{EFMeq}:

\begin{equation}
    \label{EFeqm1}
    \frac{d E(\eta)}{d\eta}=-\langle \hat{J}^1_{Q,dS} \rangle
\end{equation}
Here $\langle\hat J^1_{Q,\text{dS}}\rangle$ denotes the expectation value of the spatial current in the vacuum and is given by
\begin{equation}
\label{currentds}
    \langle \hat{J}^1_{Q,dS}\rangle = \frac{4 q^2 A_1(\eta )}{a^2(\eta)}\int_{0}^{\infty} \frac{dk}{2\pi} \langle |\phi_k|^2 \rangle
\end{equation}
Using \ref{expphi} in \ref{currentds} we have
\begin{equation}
\label{currentds1}
    \langle \hat{J}^1_{Q,dS}\rangle = \frac{ q^2 A(\eta )}{a^2(\eta)}\int_{0}^{\infty} \frac{dk}{2\pi} \frac{1}{Re(\alpha_{k,dS}(\eta))}
\end{equation}

\begin{figure}
\centering
 \includegraphics[scale=.50]{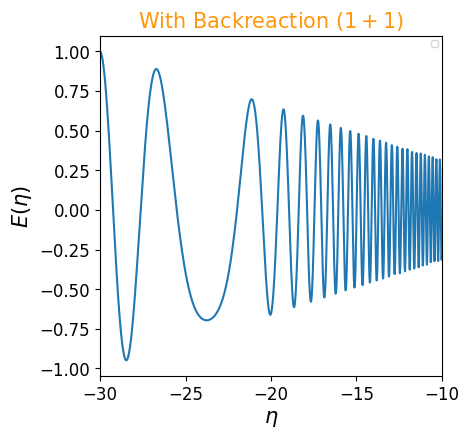}\hspace{1.0cm}
\includegraphics[scale=.50]{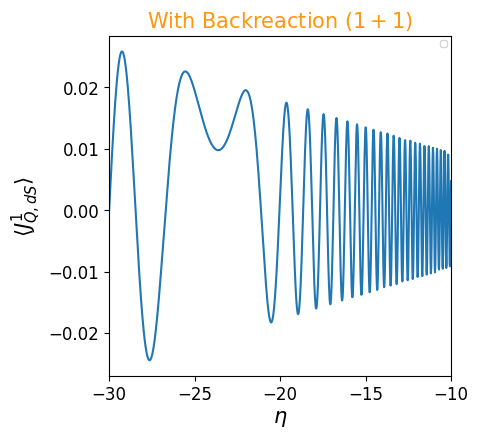}\\
    \includegraphics[scale=.50]{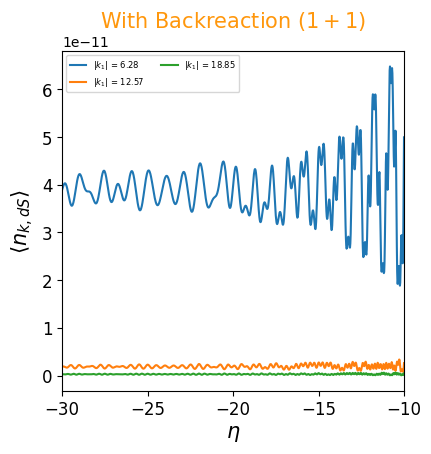}\hspace{1cm}
\includegraphics[scale=.50]{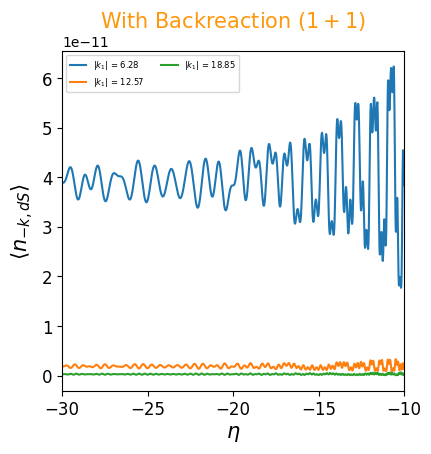}\\
\includegraphics[scale=.50]{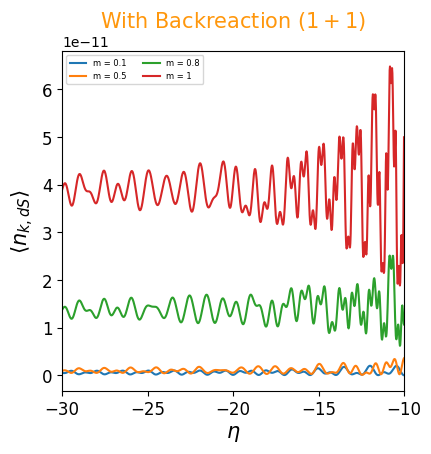}\hspace{1cm}
\includegraphics[scale=.50]{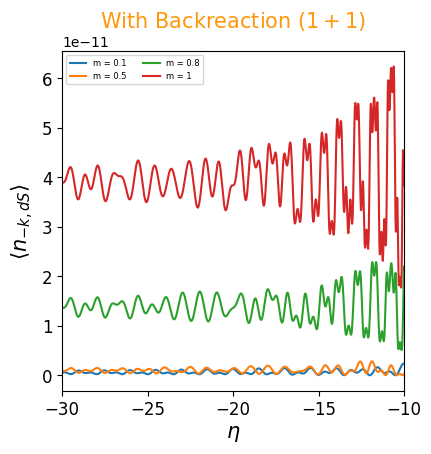}
    \caption{\small The first row shows the time evolution of the electric field $E(\eta)$ (left) and current $\langle J^1_{Q,\text{dS}} \rangle$ (right) in $(1+1)-$ dimensions de Sitter spacetime for $m = 1.0$, $E_0 = 1.0$, $q = 1.0$, and $H = 1.0$, sourced by particle-antiparticle creation. The second and third rows display $\langle n_{k,\text{dS}} \rangle$ and $\langle n_{-k,\text{dS}} \rangle$ versus $\eta$ including backreaction, showing variations with (i) different $|k_1|$ at fixed $m = 1.0$, $E_0 = 1.0$, $q = 1.0$, $H = 1.0$, and (ii) different $m$ at $|k_1| \approx 2\pi$ with other parameters fixed.
 }
    \label{fig:BRDS2}
\end{figure}

On substituting \ref{currentds1} in \ref{EFeqm1}, we have
\begin{equation}
    \label{EFeqnm1}
    E^\prime= -\frac{ q^2 A(\eta )}{a^2(\eta)}\int_{0}^{\infty} \frac{dk}{2\pi} \frac{1}{Re(\alpha_{k,dS}(\eta))}
\end{equation}
The integral in \ref{EFavg} is ultraviolet divergent; to regularize it, we replace the continuum by a one-dimensional lattice of spacing $\ell$, which yields
\begin{equation}
    \label{EFavglatticedS}
     E^\prime=-\frac{ q^2 A(\eta )}{\ell a^2(\eta)}\sum_n\frac{1}{Re(\alpha_{k,dS}(\eta))}
\end{equation}
where the sum is over all lattice points denoted by $n$. The equations \ref{aprime}, \ref{alphadedS}, and \ref{EFavglatticedS} are mutually dependent, forming a coupled system. Solving this system enables us to determine the variations in the electric field, current density $\langle J^1_{Q,dS} \rangle$, and the average number density of particles with momenta $k$ and $-k$, as functions of the parameter $\eta$, as shown in \ref{fig:BRDS2}.

\begin{figure}
\centering
     \includegraphics[scale=.54]{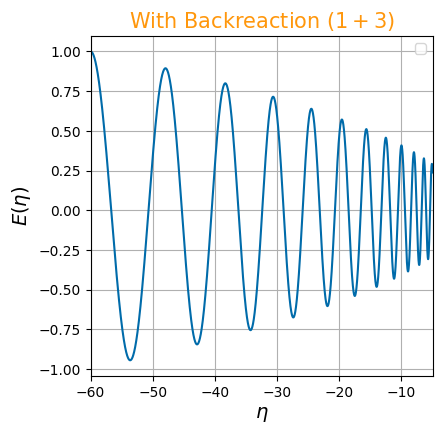}\hspace{1cm}
\includegraphics[scale=.54]{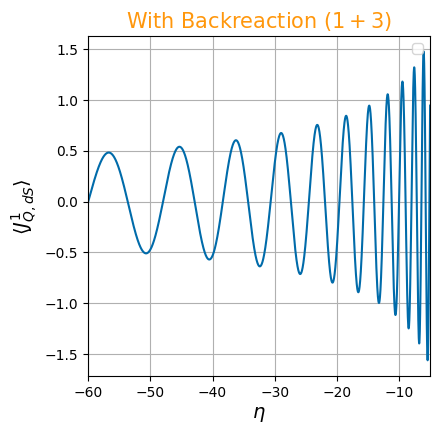}\\ 
    \includegraphics[scale=.38]{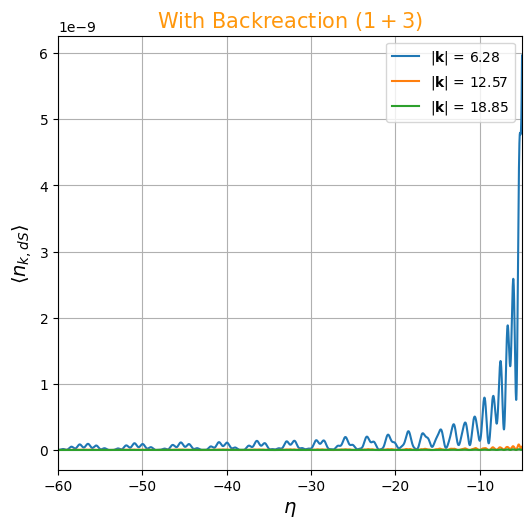}\hspace{0.1cm}
\includegraphics[scale=.38]{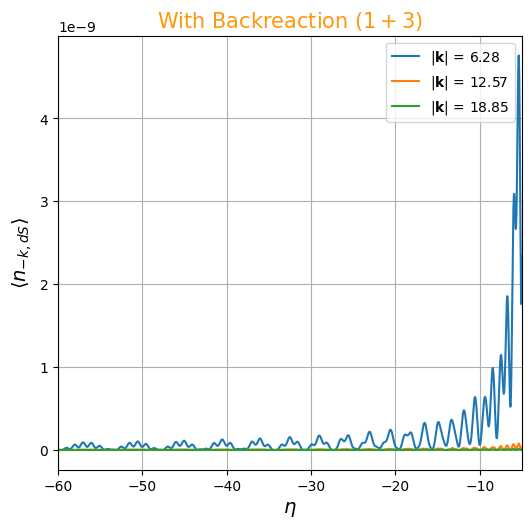}\hspace{0.1cm}
\includegraphics[scale=.38]{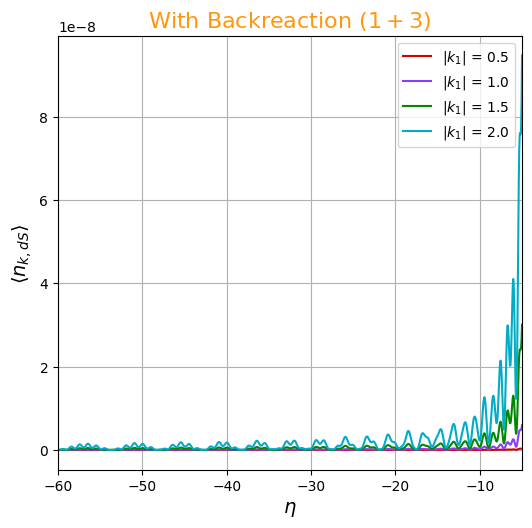}
\\\includegraphics[scale=.38]{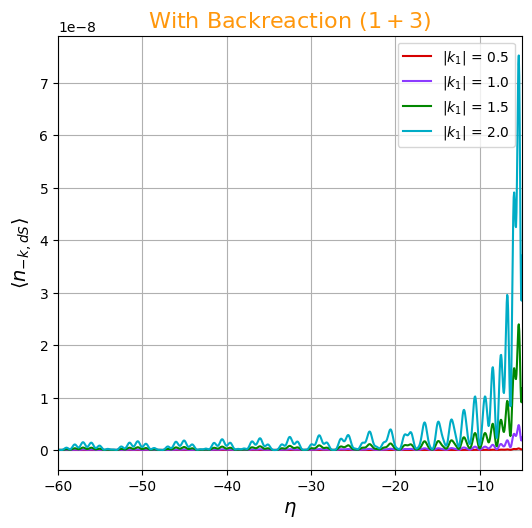} \hspace{0.1cm}
\includegraphics[scale=.38]{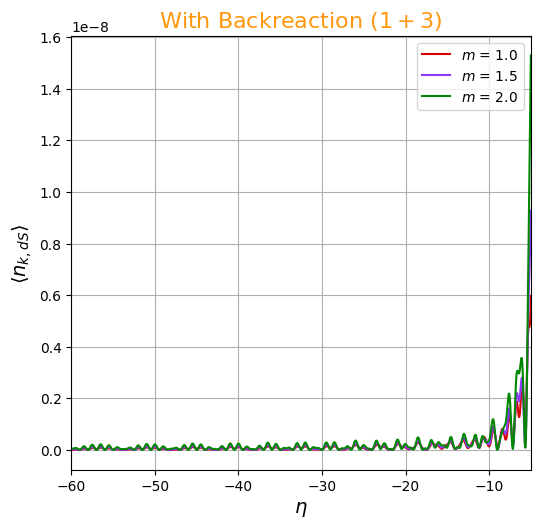} \hspace{0.1cm}
\includegraphics[scale=.38]{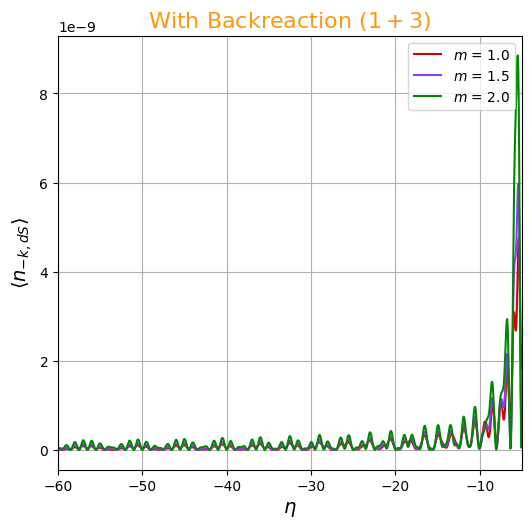}
    \caption{\small The first row shows the time evolution of the electric field $E(\eta)$ (left) and current $\langle J^1_{Q, \text{dS}} \rangle$ (right) in $(1+3)-$de Sitter spacetime, sourced by particle-antiparticle production. The second and third rows display $\langle n_{k, \text{dS}} \rangle$ and $\langle n_{-k, \text{dS}} \rangle$ versus $\eta$ including backreaction, showing variations with different $|\mathbf{k}|$ at $m=0.1$, $E_0=1$, $q=1.0$, $|k_1|=1$; different $|k_1|$ at $m=0.1$, $E_0=1$, $q=1.0$, $|\mathbf{k}|=2\pi$; and different $m$ at $|k_1|=1$, $E_0=1$, $q=1.0$, $|\mathbf{k}|=2\pi$.
}
    \label{fig:BRDS1}
\end{figure}

In de Sitter spacetime, the created particles induce plasma-like oscillations of the electric field, but unlike Minkowski space, the expansion redshifts the modes through the \(1/a^2(\eta)\) factor, causing the amplitude to decay. Quantum backreaction superimposes damped oscillations on this decay. The particle number $\langle n_{k}\rangle$ shows the same pattern as the electric field: an initial rise followed by oscillations of decreasing amplitude. In $(1+3)-$dimensions, the transverse momentum increases the effective mass, leading to a higher oscillation frequency and stronger damping than in $(1+1)$ dimensions. This underlines how cosmic expansion and dimensionality shape quantum-field dynamics in curved spacetime, unlike the undamped oscillations in Minkowski space. The corresponding effects of backreaction on the electric field, current density and particle number in $(1+3)$-dimensional de Sitter space are shown in~\ref{fig:BRDS1}.

%The number density \( \langle n_k \rangle \) reflects this behavior: it grows initially and then oscillates with decreasing amplitude. In \((1+3)-\)dimensions, transverse momentum increases the effective mass, resulting in higher oscillation frequency and stronger damping compared to \((1+1)-\)dimensions. These features highlight the crucial role of cosmic expansion and dimensionality in shaping quantum field dynamics in curved spacetime, in contrast to the undamped oscillations seen in Minkowski space.

%Furthermore, we have extended the analysis to $(1+3)$-dimensional de Sitter spacetime with backreaction, just as in the Minkowski spacetime case. The effects of backreaction on the electric field, current density, and particle number density are illustrated in ~\ref{fig:BRDS1}.

\section{Conclusion and outlook}
\label{Conclusion and outlook}

A complete quantum mechanical treatment being intractable in several realistic systems\footnote{In the case of the emergent gravity paradigm, gravity should perhaps be treated as classical throughout interacting with quantum matter.}, we consider a semiclassical description of the systems wherein certain parts of the system can be taken as classical while others to be quantum. This requires a self-consistent framework for ascertaining the evolution of the complete system. Without any guiding principles, there can be different prescriptions. The simplest procedure is to consider an effective Hamiltonian in the canonical picture, comprising of separable classical and quantum parts. One then posits a quantum evolution through the time-dependent Schr\"odinger equation for the quantum part, where the classical bit enters as a c-numbered degree of freedom, which in turn evolves through the Poisson bracket with the effective Hamiltonian, where the quantum operators can be replaced with the expectation values in the evolving state. This does achieve a self-consistent framework for the hybrid classical-quantum dynamics.

We apply this framework to the Schwinger mechanism in scalar quantum electrodynamics, examining backreaction effects of particle creation in $(1+1)$- and $(1+3)$-dimensional Minkowski and de Sitter spacetimes. Initially in ~\ref{sec:canonical}, we studied particle production without backreaction and found that the particle number density increases with $\tau$ and then saturates, remaining symmetric under $|k| \leftrightarrow -|k|$ (~\ref{velocitywbr}).

%We test out the above framework in the case of the Schwinger mechanism in scalar quantum electrodynamics by investigating the effect of backreaction of the particles created by the electric field in $(1+1)$- and $(1+3)$-dimensional Minkowski and de Sitter spacetimes. Initially, in \ref{sec:canonical} we analysed particle production due to a background electric field without backreaction in $(1+1)-$ and $(1+3)-$dimensional Minkowski spacetime. We obtain the variation of the average number of particles created with respect to $\tau$ for different $\lambda$ values, as discussed in \ref{velocitywbr}. We observed that the number density increases and then saturates, remaining symmetric under the transformation $|k| \leftrightarrow -|k|$.
We then incorporated backreaction into our framework in~\ref{Schwinger effect with backreaction}, allowing the electric field and current to evolve self-consistently with the produced particles. In Minkowski spacetime, this produced sustained plasma-like oscillations in both the electric field and current. The particle number density $\langle n_k \rangle$ likewise oscillated—remaining symmetric under $|k| \leftrightarrow -|k|$-with a constant amplitude in time. In $(1+3)-$dimensions, these oscillations became more rapid, as transverse momentum modes increased the effective frequency and strengthened the feedback.

In Sec.~\ref{Schwinger effect in the cosmological de Sitter spacetime}, we examined particle production by a background electric field in $(1+1)$- and $(1+3)$-dimensional conformally flat de Sitter spacetimes using a canonical approach. We studied how the average particle number varies with $\xi$ for different values of $M$, $L$, and $|k|$ (Fig.~\ref{DS1}). Unlike in Minkowski spacetime, cosmic expansion breaks the symmetry between $|k|$ and $-|k|$, leading to unequal production rates, which we analyzed systematically through their dependence on $M$, $L$, and the momentum modes.

%In \ref{Schwinger effect in the cosmological de Sitter spacetime}, we investigated the particle number density generated by a background electric field in $(1+1)$ and $(1+3)$-dimensional conformally flat de Sitter spacetime using a canonical approach. We analyze how the average number of particles created with momenta $|k|$ and $-|k|$ varies with the parameter $\xi$ for different values of $M$, $L$, and $|k|$, as illustrated in \ref{DS1}. Unlike the Minkowski case, in de Sitter spacetime the expansion breaks the symmetry between particles with momenta \( |k| \) and \( -|k| \), leading to unequal production rates. This asymmetry was systematically analyzed by studying the dependence of the number density on parameters such as mass \( (M) \), electric field strength \( (L) \), and momentum modes.

In Sec.~\ref{Schwinger effect with backreaction in de Sitter spacetime}, we included the backreaction of the created particles on the electric field, current, and particle number density in $(1+1)$- and $(1+3)$-dimensional de Sitter spacetimes. Backreaction produces oscillations in both the electric field and current with respect to conformal time $\eta$ (Figs.~\ref{fig:BRDS2} and \ref{fig:BRDS1}), but—unlike in Minkowski spacetime—the oscillation amplitude gradually decreases due to cosmological expansion. This damping, driven by the $1/a^{2}(\eta)$ dilution factor, redshifts field modes and suppresses the field strength over time. The particle number density $\langle n_{k}\rangle$ shows a similar pattern: initial growth followed by oscillations with diminishing amplitude. In $(1+3)-$ dimensions, transverse momentum modes further suppress particle production and enhance damping by increasing the effective mass.

%In \ref{Schwinger effect with backreaction in de Sitter spacetime}, we incorporated the backreaction of the created particles on the electric field, current, and the average number of particles with momenta \(|k|\) and \(-|k|\) in \((1+1)\)- and \((1+3)\)-dimensional de Sitter spacetimes. Our findings show that backreaction induces oscillations in both the electric field and current with respect to the conformal time \(\eta\), as shown in \ref{fig:BRDS2} and \ref{fig:BRDS1}. However, unlike in the Minkowski case, the amplitude of these oscillations gradually decreases as conformal time progresses. This damping is primarily due to the cosmological expansion, encoded in the \(1/a^2(\eta)\) dilution factor, which redshifts the field modes and naturally suppresses the field strength over time. The particle number density \(\langle n_k \rangle\) similarly exhibits initial growth followed by oscillations with diminishing amplitude. In \((1+3)\) dimensions, transverse momentum modes further suppress particle production and enhance damping due to their contribution to the effective mass.
Overall, our results underscore the effectiveness of the semiclassical approach in capturing the interplay between quantum particle production and classical field dynamics, with background geometry, dimensionality, and backreaction playing key roles. Future work will extend this framework to spacetime-dependent gauge fields (requiring finite-element methods to handle mode mixing) and explore how backreaction affects particle–antiparticle entanglement, including in the presence of a combined electric and magnetic field as in \cite{Ebadi:2015kqa, Bhattacharya:2020sjr, Ali:2021jch, Kaushal:2022las, Kaushal:2024cbk}.

%Overall, our results demonstrate the power and versatility of the semiclassical approach in capturing the interplay between quantum particle production and classical field dynamics. The contrasting behaviors in Minkowski and de Sitter spacetimes highlight the crucial role of background geometry, dimensionality, and backreaction in shaping quantum field evolution in curved spacetime.

%In future work, we aim to investigate the effects of backreaction caused by a spacetime-dependent gauge field, which will require the use of a finite element basis to address mode mixing. It would also be interesting to examine how this backreaction influences the entanglement correlations between the created particles and antiparticles and to investigate how these correlations are impacted by the addition of a background magnetic field alongside the electric field, as previously studied in \cite{Ebadi:2015kqa, Bhattacharya:2020sjr, Ali:2021jch, Kaushal:2022las, Kaushal:2024cbk} with no backreaction. 

Further, there exists another prescription of ``A healthier semi-classical dynamics" \cite{SC4} which claims to offer a better handle on the self-consistency through linear dynamics in the quantum-classical state. This new scheme could be better, as the simplest scheme used here that uses the expectation values in the effective Hamiltonian did have issues in a toy model of a harmonic oscillator coupled with two qubits \cite{Husain:2022kaz}. It is pertinent to develop the investigate the new recipe in the scalar quantum electrodynamics and beyond for any discernible differences.

\bibliographystyle{cas-model2-names}

%\bibliography{cas-refs}

%%%%%%%%%%%%%%%%%%%%%%%%%%%%%%%%%%%%%%
%\section{References}
%%%%%%%%%%%%%%%%%%%%%%%%%%%%%%%%%%%%%

\end{document}